\documentclass[11pt,twoside]{article}

\usepackage{asp2004}
\usepackage{epsf}
\usepackage{psfig}
\usepackage{lscape}
\usepackage{graphicx}
\usepackage{subfigure}

\begin{document}
\title{The Mass Distribution of White Dwarfs: An Unwavering Obsession}   
\author{P. Bergeron$^1$, A. Gianninas$^1$, and S. Boudreault$^2$}   
\affil{$^1$D\'epartement de Physique, Universit\'e de Montr\'eal, C.P. 6128, 
Succ. Centre-Ville, Montr\'eal, Qu\'ebec, Canada, H3C 3J7}
\affil{$^2$Max-Planck-Institut f\"ur Astronomie, K\"onigstuhl 17, D-69117 Heidelberg, Germany} 

\begin{abstract} 
We discuss some of our current knowledge of the mass distribution
of DA and non-DA stars using various methods for measuring white dwarf
masses including spectroscopic, trigonometric parallax, and
gravitational redshift measurements, with a particular emphasis on the
problems encountered at the low end of the cooling sequence where
energy transport by convection becomes important.
\end{abstract}



\section{An Unwavering Obsession}

The unwavering obsession to which the title refers applies only to the
first author since the other co-authors are still too young to be
obsessed by such a thing as the mass distribution of white dwarf
stars. 

As early as 1976, it was suggested that below $T_{\rm eff}\sim
12,000$~K, convective mixing between the thin superficial hydrogen
layer and the more massive underlying helium layer could turn a
hydrogen-rich star into a helium-rich star, provided the mass of the
hydrogen layer is small enough (a modern value yields $M_{\rm H}\simeq
10^{-6}\ M_\odot$). Furthermore, the effective temperature at which
this mixing occurs is a function of the mass of the hydrogen layer:
for thicker hydrogen layers, the mixing occurs at lower effective
temperatures. Since the process of convective mixing is still poorly
understood, the exact ratio of helium to hydrogen after the mixing
occurs remains unknown. In particular, it is possible that instead of
turning a DA star into a featureless helium-rich DC star, convective
mixing may simply enrich the hydrogen-rich atmosphere with large
quantities of helium, leading to a mixed hydrogen and helium
atmospheric composition. Such a hypothesis is difficult to test,
however, since helium becomes spectroscopically invisible below
$T_{\rm eff}\sim12,000$~K, and its presence can only be inferred
through indirect methods.

Such a method has been proposed by \citet{LW83} who showed that the
atmospheric helium abundance could be determined from a detailed
examination of the high Balmer lines, since the presence of helium
increases the photospheric pressure, and thus produces a quenching of
the upper levels of the hydrogen atom which, in turn, affects the line
profiles. This method has been put forward on a more quantitative
basis by
\citet{bergeron90} who analyzed 37 cool DA stars using the spectroscopic
method of fitting high Balmer line spectroscopy with the predictions
of detailed model atmospheres with mixed hydrogen and helium
compositions.  Their analysis first showed that the effects produced on
the hydrogen lines at high $\log g$ could not be distinguished
from those produced by the presence of large amounts of helium. Hence,
the problem could only be approached from a statistical point of view
by assuming a mean value of $\log g=8$ for all stars, and then by
determining individual helium abundances. Under this assumption, the
analysis of Bergeron et al.~revealed that the atmospheres of most
objects below $T_{\rm eff}\sim 11,500$~K were contaminated by significant
amounts of helium, with abundances sometimes as high as $N({\rm He})/N({\rm H})
\sim 20$. 

We show in Figure 1 an update of this result using a sample of 232 DA
stars analyzed with our most recent grid of model atmospheres. On the
left panel we show the surface gravity as a function of effective
temperature for each object. Clearly, the values determined here are
significantly higher than the canonical value of $\log g=8$ for DA
stars (shown by the dashed line); the mean surface gravity of this
sample is actually $\log g=8.167$. If we assume instead that our
sample is representative of other DA stars and adopt $\log g=8$ for
each object, we can determine individual helium abundances. This is
shown on the right panel of Figure 1. As can be seen, non-negligible
amounts of helium in the range $N({\rm He})/N({\rm H})=0.1-10$ at the
surface of these DA stars can easily account for the high $\log g$
values inferred under the assumption of pure hydrogen compositions.

\setcounter{figure}{0} 
\begin{figure}[h] 
\begin{center}
\includegraphics[trim=100 200 100 200,height=6.5cm,angle=270]{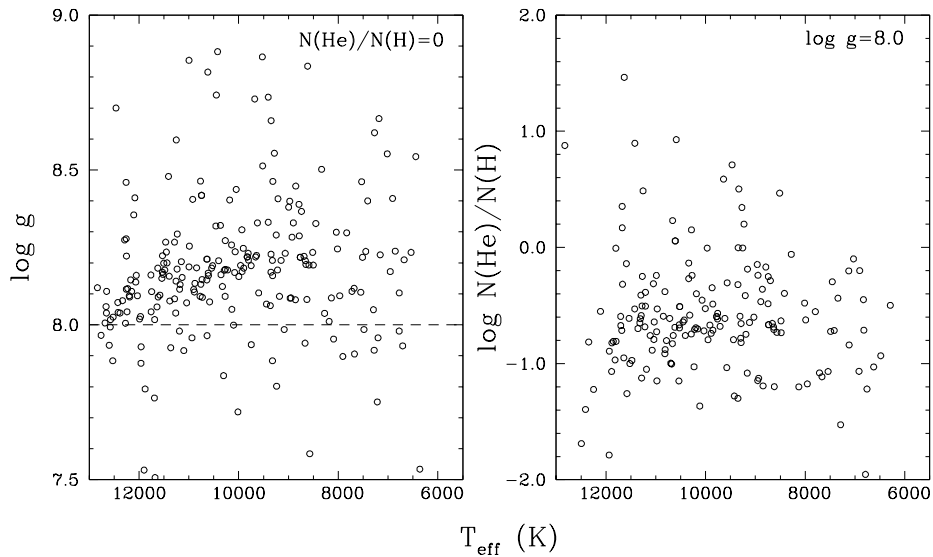}
\end{center}
\caption{Atmospheric parameters for a sample of 232 DA stars obtained under
the assumption of a pure hydrogen atmospheric composition (left panel)
and under the assumption of an average surface gravity of $\log g=8.0$
(right panel).}
\end{figure}

\section{The Spectroscopic Method at High Effective Temperatures}

The results discussed above rest heavily on the abililty of the models
to describe accurately the physical conditions encountered in cool
white dwarf atmospheres, but also on the reliability of the
spectroscopic method to yield accurate measurements of the atmospheric
parameters. It is with this idea in mind that \citet[][BSL
hereafter]{BSL} decided to test the spectroscopic method using
DA white dwarfs at higher effective temperatures ($T_{\rm
eff}>13,000$~K) where the atmospheres are purely radiative and thus do
not suffer from the uncertainties related to the treatment of
convective energy transport, and where the assumption of a pure
hydrogen composition is certainly justified.  From the analysis of a
sample of 129 DA stars, BSL determined a mean surface gravity of $\log
g=7.909$, in much
better agreement with the canonical value of $\log g=8$ for DA stars.

\begin{figure}[h] 
\begin{center}
\includegraphics[trim=100 120 120 120,height=8.cm,angle=0]{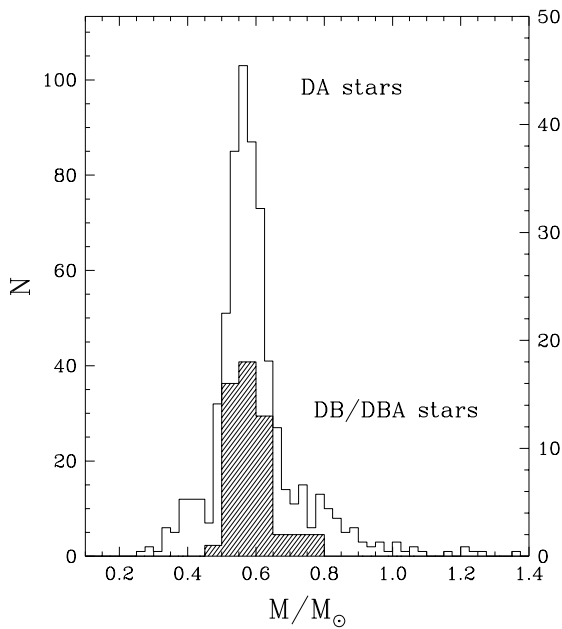}
\end{center}
\caption{Mass distribution of 677 DA stars above $T_{\rm eff}=13,000$~K 
(solid line; left axis) compared with that of 54 DB and DBA stars
above $T_{\rm eff}=15,000$~K (hatched histogram; right axis). The
average masses are 0.585 and 0.598 $M_\odot$, respectively.}
\end{figure}

More recently, \citet{LBH} obtained high signal-to-noise spectroscopy
of all 348 DA stars from the Palomar Green Survey and determined the
atmospheric parameters for each object using NLTE model
atmospheres. If we restrict the range of effective temperature to
$T_{\rm eff}>13,000$~K, the mean surface gravity of their sample is
$\log g=7.883$, in excellent agreement with the results of BSL. The
corresponding mean mass for this sample is $0.603\ M_\odot$ using
evolutionary models with thick hydrogen layers. As part of our ongoing
survey aimed at defining more accurately the empirical boundaries of
the instability strip (see Gianninas, Bergeron, \& Fontaine, these
proceedings), we have been gathering for several years optical
spectroscopy of DA white dwarfs from the McCook \& Sion catalog using
the Steward Observatory 2.3 m telescope facility. The mass
distribution for the 667 DA stars above 13,000 K is displayed in
Figure 2, together with the mass distribution for 54 DB and DBA stars
taken from
\citet{beauchamp96}; for the latter, uncertainties with the line broadening 
theory of helium lines limits the accuracy of the spectroscopic method
to $T_{\rm eff}>15,000$~K. Both mass distributions are in excellent
agreement.

\section{Moving Towards Lower Effective Temperatures}

The results discussed in the last section indicate that the
atmospheric parameters of hot ($T_{\rm eff}>13,000$~K) DA stars are
reasonable, and that the high $\log g$ values obtained for cool DA
stars are not related directly to the spectroscopic method itself. One
of the largest uncertainties in cool white dwarf atmospheres is the
treatment of convective energy transport. In an effort to parameterize
the convective efficiency in DA stars,
\citet{bergeron95} studied a sample of 22 ZZ Ceti stars and showed
that the so-called ML2/$\alpha=0.6$ version of the mixing-length
theory yields the best internal consistency between optical and UV
effective temperatures, trigonometric parallaxes, $V$ magnitudes, and
gravitational redshift measurements. The mass vs.~temperature
distribution for our complete sample of DA stars using this
parameterization is displayed in Figure 3.

\begin{figure}[h] 
\begin{center}
\includegraphics[trim=100 220 120 230,height=5.5cm,angle=270]{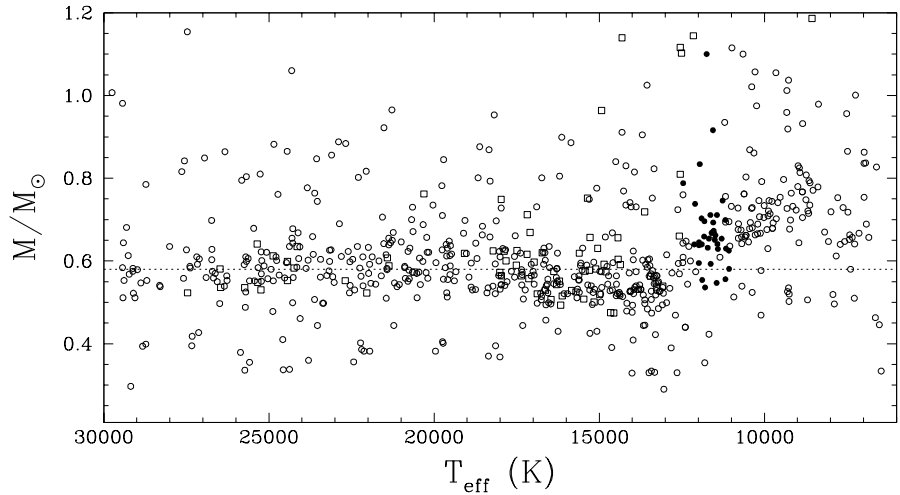}
\end{center}
\caption{Spectroscopic masses as a function of effective temperature for our
complete sample of DA stars (open circles) and DB/DBA stars
(open squares). Filled circles correspond to ZZ Ceti stars.}
\end{figure}

The problem with the high masses --- or high $\log g$ values --- towards low
effective temperatures is clearly apparent here. However, this increase 
in mass begins not only where
convective mixing is believed to occur, but even in the temperature 
range where ZZ Ceti stars are found. If
the larger-than-average mass for ZZ Ceti stars is explained in terms
of helium enrichment from the deeper helium convection zone, the
hydrogen layers in ZZ Ceti stars need to be as thin as $M_{\rm H}\sim
10^{-12}\ M_\odot$ (see Fig.~4 of Dufour \& Bergeron, these
proceedings), in sharp contrast with values determined for ZZ Ceti
stars from asteroseismology. 

\section{Alternative Means of Measuring White Dwarf Masses}

Further insight into the problem with high masses at low effective
temperatures may be gained by comparing the masses inferred from the
spectroscopic method with those obtained from other methods, namely
from trigonometric parallax and gravitational redshift measurements.
We have thus secured high signal-to-noise spectroscopy for 129 DA
stars, 92 of which have parallaxes available, and 49 of which have
gravitational redshifts. In Figure 4 we reproduce the
spectroscopic masses as a function of effective temperature for our
complete sample of DA stars discussed in the previous section, and we
overplot in the top and bottom panels the masses derived from
trigonometric parallax and gravitational redshift measurements,
respectively.

\begin{figure}[h] 
\begin{center}
\includegraphics[trim=100 120 120 160,height=9.0cm,angle=0]{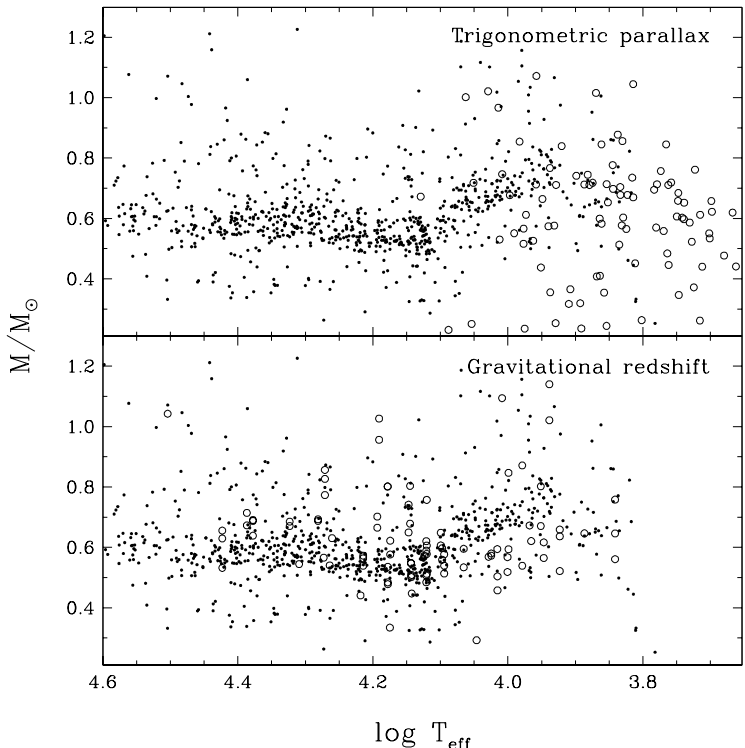}
\end{center}
\caption{Spectroscopic masses of DA stars (solid dots) compared with
masses derived from trigonometric parallax measurements (open circles,
top panel) and from gravitational redshift parallax measurements (open
circles, bottom panel). All three samples correspond to a different set stars.}
\end{figure}

The parallax method relies on optical $BVRI$ and infrared $JHK$
photometric measurements to constrain the effective temperature and
the solid angle $\pi(R/D)^2$ between the flux received at Earth and
that emitted at the surface of the star. Given the distance $D$
obtained from the parallax, we obtain directly the radius $R$ and thus
the photometric mass using evolutionary models. One of the most
obvious features of the distribution of photometric masses in the top
panel of Figure 4 is the large number of low mass ($M<0.4\ M_\odot$)
white dwarfs.  Most of these objects are probably unresolved double
degenerates for which the flux received at Earth corresponds to the
contribution of both stars. Hence the radii are overestimated and the
masses are underestimated. If we ignore these low mass objects, the
photometric masses do exhibit higher than average masses when compared
with the spectrocopic masses at higher temperatures, which are closer
to the canonical value of $\sim 0.6\ M_\odot$, although the dispersion
of the photometric masses is also significantly larger.  In some, but
not all cases, the photometric mass exceeds the spectrocopic mass.  It
is then possible to adjust individually the atmospheric helium
abundances until both methods yield consistent masses.  Since the
parallax method uses broadband energy distributions that
are not affected significantly by the presence of helium \citep[][Fig.~4]{boudreault05}, the
photometric masses are almost completely independent of the assumed
atmospheric composition, in sharp contrast with the spectroscopic
method. We finally note that in the case of the massive ($M=1.3\
M_\odot$) DA star LHS 4033 \citep{dahn04}, the parallax and
spectrosocpic masses agree to within 0.01 $M_\odot$.

The gravitational redshift method used in the bottom panel of Figure 4
is based on the relation between the measured redshift velocity, the
mass $M$, and the radius $R$ of the star, $v_{\rm GR}=GM/Rc$. This
method provides mass measurements that are almost completely
independent of anything else, although the velocity measurements are
intrinsically more difficult to obtain than the other types of
measurements used with other techniques. In particular, since the
redshift mass measurements scale linearly with velocity, low mass
white dwarfs are intrisically more difficult to measure, and the
errors are correspondingly larger. Nevertheless, if we take at face
value the results shown in Figure 4, the redshift masses do not reveal
any particular trend at low effective temperatures. More accurate
trigonometric parallax and gravitational redshift measurements of {\it
individual} stars may be required, together with some unwavering
obsession, to help us further understand the high mass problem
observed at low temperatures.

\acknowledgements 
The authors acknowledge the support of the NSERC Canada and the Royal
Astronomical Society. P. Bergeron is a Cottrell Scholar of Research
Corporation.


\begin{thebibliography}{}

\bibitem[Beauchamp et al.(1996)]{beauchamp96} Beauchamp, A., Wesemael, F., Bergeron, P., Liebert, J., \& Saffer, R. A. 1996, in ASP Conf. Ser. Vol. 96, Hydrogen-Deficient Stars, ed. S. Jeffery \& U. Heber (San Francisco: ASP), p.~295

\bibitem[Bergeron, Saffer, \& Liebert(1992)]{BSL} Bergeron, P., Saffer, R. A., \& Liebert, J. 1992, \apj, 394, 228 (BSL)

\bibitem[Bergeron et al.(1990)]{bergeron90} Bergeron, P., Wesemael, F., Fontaine, G., \& Liebert, J. 1990, \apj, 351, L21

\bibitem[Bergeron et al.(1995)]{bergeron95} Bergeron, P., Wesemael, F.,
Lamontagne, R., Fontaine, G., Saffer, R. A., \& Allard, N. F. 1995,
\apj, 449, 258

\bibitem[Boudreault \& Bergeron(2005)]{boudreault05} Boudreault, S., \& Bergeron, P.
2005, in 14th European Workshop on White Dwarfs, ASP Conf.~Series,
vol.~334, eds. D.~Koester \& S.~Moehler, p.~249

\bibitem[Dahn et al.(2004)]{dahn04} Dahn, C. C., Bergeron, P., Liebert, J., 
Harris, H. C., Canzian, B., Leggett, S. K., \& Boudreault, S. 
2004, \apj, 605, 400

\bibitem[Liebert, Bergeron, \& Holberg(2005)]{LBH} Liebert, J., Bergeron, P., \& Holberg, J. B. 2005, \apjs, 156, 47

\bibitem[Liebert \& Wehrse(1983)]{LW83} Liebert, J., \& Wehrse R. 1983, \aap, 122, 297

\end{thebibliography}
\end{document}